\documentclass[jcp,preprint,
unsortedaddress
]{revtex4-1}

\usepackage{graphicx}
\usepackage{dcolumn}
\usepackage{bm}

\begin{document}

\title{Grand-Canonical Quantized Liquid Density-Functional Theory in a Car-Parrinello Implementation}

\author{Christian~F.~J.~Walther}
 \email{c.walther@jacobs-university.de} 
 \affiliation{School of Engineering and Science, Jacobs University Bremen, Campus Ring 1, 28759 Bremen, Germany}

\author{Serguei Patchkovskii}
 \affiliation{Steacie Institute for Molecular Sciences, NRC, 100 Sussex Drive, Ottawa, Ontario, Canada K1A 0R6}

\author{Thomas~Heine}
 \email{t.heine@jacobs-university.de} 
 \affiliation{School of Engineering and Science, Jacobs University Bremen, Campus Ring 1, 28759 Bremen, Germany}

\begin{abstract}
 Quantized Liquid Density-Functional Theory [Phys. Rev. E 2009, 80, 031603], a method developed to 
 assess the adsorption of gas molecules in porous nanomaterials, is reformulated within the grand canonical 
 ensemble. With the grand potential it is possible to compare directly external and internal thermodynamic 
 quantities. In our new implementation, the grand potential is minimized utilizing the 
 Car-Parrinello approach and gives, in particular for low temperature simulations, a significant 
 computational advantage over the original canonical approaches. The method is validated against original QLDFT, 
 and applied to model potentials and graphite slit pores.
\end{abstract}

 \maketitle

\section{Introduction}
In the recent past, it has been shown that metal-organic frameworks (MOFs) and other molecular framework materials are suitable to adsorb hydrogen 
in appreciable quantities \cite{h2mof} and to separate light-weight fluids, as for example H$_2$ and D$_2$.\cite{h2d2-japanese-jacs,Teufel2013,cai2012,liu2012}
At low temperatures, quantum effects of the adsorbed fluids in the rather weak host-guest potential become important, and in the case of quantum sieving 
they determine the performance of the functional material. \cite{Beenakker1995,Teufel2013} Quantum effects are also required for the quantitative estimation 
of the gas adsorption capacity of materials, in particular towards light-weight molecules such as H$_2$ and D$_2$.\cite{Garberoglio2005,Kumar2006} At present, two principal models are established for the treatment of quantum effects: widely spread is the Feynman-Hibbs correction of the classical potential in the Grand Canonical Monte Carlo (GCMC) method\cite{FeynmanHibbs,FeynmanSM}. A more rigorous, albeit computationally more demanding, model 
is the path integral (PI) technique \cite{PI1948,FeynmanHibbs}, that has been applied to describe H$_2$ and D$_2$ clusters \cite{chakravarty1995} 
and to assess hydrogen adsorption for carbon model pores \cite{johnsonh2carbonfoam}. For a review about the path integral technique we refer to 
the works of Ceperley\cite{ceperley1995} or Chakravarty.\cite{chakravarty1997}.

In 2005, Patchkovskii et al. suggested to assess the quantum effects of light-weight molecules or atoms, i.e. of hydrogen molecules, within the ideal gas approximation by solving the Schr\"odinger equation (SE) of a single H$_2$ 
particle in the external potential.\cite{PNAS2005} The wave function of the adsorbed hydrogen fluid has been expanded in a plane wave series, and the SE has 
been solved by explicit diagonalization of the Hamiltonian. In 2007, the numerical implementation has been rigorously revised by using an exponential series 
to compute the partition function, taking advantage of the sparseness of the Hamiltonian, leading to significant computational advantage for temperatures of 200~K or higher.\cite{patchkovskii2007} In order to go beyond the ideal gas treatment, 
Patchkovskii and Heine applied the Kohn-Sham approach \cite{KS1965} to the problem of a quantum fluid adsorbed in an external potential in the so-called Quantized 
Liquid Density-Functional Theory (QLDFT).\cite{QLDFT2009} Like in the traditional Kohn-Sham method for electronic systems, the fully interacting system is mapped 
to a non-interacting system in an effective potential, and the effective one-particle states are occupied within the classical Maxwell-Boltzmann statistics.  All terms of the free energy which cannot 
be expressed explicitly as functional of the density are treated approximately in the excess functional $F_{xc}$, that has been determined such that the experimental 
isotherms of the uniform fluid served as reference system, similar as the homogeneous electron gas in electronic DFT. 

In 2011, Mesa et al. have implemented the proper Bose-Einstein statistics to QLDFT. However, they have shown that even at temperatures of 50~K the classical 
Maxwell-Boltzmann statistics does not deviate from its rigorous counterpart. On the other hand, quantum effects have been found to be appreciable even at ambient 
temperature for certain pore sizes.\cite{Mesa2011}

The QLDFT method has been successfully applied to describe hydrogen adsorption in carbon foams\cite{Mesa2012}, in Porous Aromatic Frameworks (PAFs)\cite{Lukose2012} 
and to determine the selectivity of $D_2$ vs. $H_2$ adsorption in the metal-organic framework MFU-4.\cite{Teufel2013}

Today it is generally accepted that reversible hydrogen storage by physisorption in MOFs and other nanoporous materials is only efficient at 
temperatures that allow cooling with liquid nitrogen (77~K). The presently available QLDFT code \cite{QLDFT2009} shows a rather poor computational performance 
for temperatures below 100~K. Moreover, it is advantageous to reformulate canonical QLDFT (C-QLDFT) within the grand canonical ensemble to allow a direct comparison to 
thermodynamic data through the chemical potential, and to be able to compare our simulations with GCMC calculations.

In this article we present the reformulation of QLDFT in the grand-canonical ensemble (GC-QLDFT). For computational efficiency, we suggest a 
simple approximation of the free energy that is advantageous in particular for low temperature applications. We minimize the grand potential employing 
the Car-Parrinello (CP) method \cite{Car1985} and compare the performance of this new approach for model potentials and carbon slit pores with C-QLDFT and literature data.
\section{Method}
\label{Sec:Method}
According to Hohenberg-Kohn and Mermin\cite{Hohenberg1964, Mermin1965} the exact 
grand potential $\Omega$ is a functional of the density $\rho(\vec{r})$. This functional 
can be written as 
\begin{eqnarray}
 \Omega[\rho]=F_{HKM}[\rho]-\mu\int d^3r\,\rho(\vec{r})+\int d^3r\,\rho(\vec{r})v_{ext}(\vec{r}).  
\end{eqnarray} 
$F_{HKM}$ denotes the Hohenberg-Kohn and Mermin free energy functional, $\mu$ the chemical potential 
and $v_{ext}$ is the external potential.

The physical density of a system in contact with a particle reservoir 
is obtained by fixing the chemical potential and minimizing the 
functional $\Omega$. The only problem is that the general form of the 
free energy functional $F_{HKM}$ is unknown and approximations are necessary. 
\subsection{The QLDFT free energy}
In C-QLDFT the free energy $F$ is obtained from the following expression
\begin{eqnarray}
 \label{eq: free_energy_original_QLDFT} 
 F[\rho]=&&F_{0}[\rho]-F_{H}[\rho]-\int\,d^3r\,v_{xc}(\vec{r})\rho(\vec{r})
+
F_{xc}[\rho].
\end{eqnarray} 
$F_{0}$ describes the free energy of a non-interacting system in an effective potential 
and is discussed below in more detail. $F_{xc}$ denotes the excess functional and 
$v_{xc}$ is the excess potential which can be obtained by the functional derivative 
of $F_{xc}$ with respect to the density $\rho$. $F_{H}[\rho]$ is given by
\begin{eqnarray} 
 \label{eq: hartree} 
 F_{H}[\rho]=\frac{1}{2}\int d^3r\int d^3r'\,\rho(\vec{r})v_{12}(|\vec{r}-\vec{r}'|)\rho(\vec{r}')
\end{eqnarray} 
where $v_{12}$ denotes the two-body interaction potential between the guest molecules. The choice of $v_{12}$ may range from complete neglection of this typically slightly attractive term to parametrized data from {\em ab initio} calculations. For hydrogen adsorption, we have chosen to employ the zero-order term of the extrapolated exact Born-Oppenheimer intermolecular H$_2$--H$_2$  interaction potential, as reported by Diep and Johnson.\cite{diep2000} 
$F_{0}$ is calculated from the non-interacting one-particle partition function $Z_{0}$ 
using 
\begin{equation}
\label{eq:F0Old}
F_{0}=-\frac{N}{\beta}ln\,\frac{Z_{0}}{N}. 
\end{equation}
$Z_{0}$ 
is given by a power series 
expansion of the non-interacting effective Hamiltonian $\mathbf{H_{0}}$ 
\begin{eqnarray} 
 \label{eq: h0} 
 \mathbf{H_{0}}=&&-\frac{\vec{\nabla}^2}{2m}+v_{ext}(\vec{r})
 +\int d^3r'\,v_{12}(|\vec{r}-\vec{r}'|)\rho(\vec{r}')
 +v_{xc}(\vec{r})
\end{eqnarray}
of $N$ particles in the external host--guest potential $v_{ext}$: 
\begin{eqnarray} 
 Z_{0}=\mbox{Tr}\,\exp(-\beta\mathbf{H_{0}})
 \exp(-\beta\mathbf{H_{0}})\approx\mathbf{1}+\sum_{n=1}^{n_{max}}\frac{1}{n!}(\mathbf{-\beta H_{0}})^{n}.
\end{eqnarray} 
with $\beta=(k_B T)^{-1}$. The numerical effort to treat the power series depends on the temperature. In case 
of high temperatures convergence is rapidly achieved for small numbers of $n_{max}$, but for low temperatures, i.e. at 77~K, 
convergence requires the inclusion of a large number of terms. Therefore, a simplification 
to avoid the power series is appreciated and will be suggested below.
The power series was originally introduced because of the mapping of an interacting system 
onto a system of non-interacting particles in an effective potential. In principle, 
this mapping is very useful because one can obtain a very good approximation of the free 
energy. It also yields one-particle wave functions in that formalism, and the explicit implementation of 
the Bose-Einstein statistics is straight-forward.\cite{Mesa2011} However, for lower temperatures this approach becomes computationally prohibitively expensive and therefore requires revision.
\subsection{The GC-QLDFT grand potential}
The C-QLDFT free energy functional will be reformulated in this subsection, avoiding the power series that has marked the 
computational bottleneck in C-QLDFT. We will obtain the GC-QLDFT grand potential which is
based on the minimum principle.
The free energy of the effective one-particle reference system $F_{0}$ can be written as follows

\begin{eqnarray}
 \label{eq:F0}
 F_{0}[\rho]&&=-\frac{N}{\beta}ln\,\frac{Z_{0}}{N} 
 =-\frac{N}{\beta}\left(ln\,Z_{0}-ln\,N\right)\nonumber\\
 &&=N\left(\sum_{i}f_{i}\int d^3r \varphi_{i}(\vec{r})^{*}\mathbf{H_{0}}\varphi_{i}(\vec{r})
 +\frac{1}{\beta}\sum_{i}f_{i}\log(f_{i})\right)
 +\frac{N}{\beta}ln\,N.
\end{eqnarray} 
$\varphi_{i}(\vec{r})$ are eigenfunctions of the effective Hamiltonian $\mathbf{H_{0}}$ 
and $f_{i}$ can be obtained from the eigenvalues by using $Z_0^{-1} \exp[-\beta\varepsilon_i]$.
In C-QLDFT the density $\rho(\vec{r})$ is calculated by rescaling the non-interacting 
density matrix 
%
\begin{eqnarray}
 \gamma_{0}(\vec{r},\vec{r}')=\sum_{i}\varphi_{i}(\vec{r})^{*}f_{i}\varphi_{i}(\vec{r}),
\end{eqnarray}
yielding a fluid density 
\begin{eqnarray}
 \rho(\vec{r})=N\gamma_{0}(\vec{r},\vec{r}).
\end{eqnarray} 
We insert the definition of the Hamiltonian $\mathbf{H_{0}}$ given in equation \ref{eq: h0} into 
\ref{eq:F0}. By using equation \ref{eq: free_energy_original_QLDFT} and some simplifications we can write 
the grand potential in the minimum with $n_{i}:=Nf_{i}$ in the following way
\begin{eqnarray}
 \label{eq:CPGCQLDFTOMEGA}
 \Omega_{0}&&=\min_{n_{i},\varphi_{i}}\Big[\sum_{i}n_{i}\int d^3r\,\varphi_{i}(\vec{r})^{*}\Big(-\frac{\vec{\nabla}^2}{2m}\Big)\varphi_{i}(\vec{r})\nonumber\\
&&+\int d^3r\,\rho(\vec{r})v_{ext}(\vec{r})+\frac{1}{\beta}\sum_{i}n_{i}\log(n_{i})\nonumber\\
&&+F_{H}[\rho]+F_{xc}[\rho]-\mu\int d^3r\,\rho(\vec{r})\nonumber\\
&&+\sum_{i,j}\lambda_{i,j}\left(\int d^3r \varphi_{i}^{*}(\vec{r})\varphi_{j}(\vec{r})-\delta_{i,j}\right)\Big].
\end{eqnarray}  
$\lambda_{i,j}$ are Lagrange multipliers which were introduced to keep the orthogonality of the set 
of eigenfunctions $\varphi_{i}$. 
\subsection{The zero-temperature reference system}
To simplify equation \ref{eq:CPGCQLDFTOMEGA} we apply the zero-temperature 
approximation to the one particle system, but keep the finite temperature in 
the many particle system. Such a reference system would be inappropriate  
if we would describe electrons, since the electron-electron interaction is quite 
soft and therefore the spin statistics is important in establishing the pair-correlation 
function. 
But in case 
of molecules the intermolecular interaction gets very hard at short distances and completely 
overwhelms the spin statistics in the pair-correlation function. The work of Mesa et al. 
\cite{Mesa2011} has already shown that Maxwell-Boltzmann and Bose-Einstein statistics are equivalent even for temperatures as low as 50~K. 

As the zero temperature reference system appears to be a rather strong approximation we have carefully validated it, and we will 
show that it yields density profiles in excellent agreement with the C-QLDFT method.
     
For the zero-temperature approximation and a real wave function we 
obtain the following expression for the free energy
\begin{eqnarray}
  \label{eq:qLIE1_1}
  F[\rho]&&=-\int d^3r \phi(\vec{r})\frac{\vec{\nabla}^2}{2m}\phi(\vec{r})
  +F_{H}[\rho]
  +F_{xc}[\rho]+\int d^3r\,\rho(\vec{r})v_{ext}(\vec{r}).
\end{eqnarray} 
Here we have defined $\phi(\vec{r}):=\sqrt{N}\varphi_{0}(\vec{r})$. Note that 
with such choice of the free energy the grand potential $\Omega$ can be 
minimized without any constraint.
%

Different suggestions for the choice of $F_{xc}$ 
have already been discussed in ref. \cite{QLDFT2009}. There, the most sophisticated functional is the LIE1 
(LIE=local interaction expression) excess functional which uses the weighted local 
density approximation (WLDA) \cite{WLDA1985}. This functional is based on the 
parametrized thermodynamic data summarized by McCarty et al.\cite{mccarty1981} and 
based on the experiment of Mills et al .\cite{mills1977}. A double counting term is subtracted. 
$F^{LIE1}[\rho]$ is given by
\begin{eqnarray}
 &&F^{LIE1}[\rho]=\int d^3r\,\rho(\vec{r})\Big\{f_{exp}[\rho(\vec{r})]
 +\beta^{-1}\log\big(\frac{Z_{kin}}{\overline{\rho}(\vec{r})V}\big)-\frac{1}{2}\rho(\vec{r})\int d^3r' v_{12}(\vec{r}')\Big\}, 
\end{eqnarray}          
where $Z_{kin}$ is the kinetic partition function obtained from the cyclic-Hamiltonian limit. $f_{exp}$ denotes the 
free energy per particle obtained in experiment and the weighted density is given by 
\begin{eqnarray}
 \overline{\rho}(\vec{r})=\int d^3 r' w(|\vec{r}-\vec{r}'|)\rho(\vec{r}').
\end{eqnarray} 
$Z_{kin}$, $\overline{\rho}$ and $f_{exp}$ are utilized in exactly the same way as in C-QLDFT and for details we refer to ref. \cite{QLDFT2009}.
We recognized that $F_{xc}$ cannot simply be replaced by $F_{xc}^{LIE1}$ as the 
zero temperature approximation is applied to the one-particle system. This approximation affects 
the kinetic energy of the effective one particle system. The kinetic part of the 
double counting term (the term $\beta^{-1}\log\big(\frac{Z_{kin}}{\overline{\rho}(\vec{r})V}\big)$) 
is in principle not necessary in case of the zero temperature approximation but it controls the 
density oscillations within the WDA and can therefore not be omitted. 
For systems involving appreciable density oscillations we
found that 
\begin{eqnarray}
 \label{eq:qLIE1_2}
 F_{xc}=F_{xc}^{LIE1}-\beta^{-1}\int d^3r\,\rho(\vec{r})\log\left(\frac{Z_{kin}}{\rho(\vec{r})V}\right)
\end{eqnarray} 
yields satisfactory density profiles for the model systems we have studied. We will call this 
functional qLIE1 functional (quantum LIE1).
\subsection{Approximation of $F_{HKM}$}
With the formalism suggested in this work it is possible to approximate the 
Hohenberg-Kohn and Mermin free energy $F_{HKM}$ similar to the strategy used in classical 
liquid density functional theory (CLDFT) \cite{saam1977,evans1979,singh1991}.  

The most straight-forward approximation of the free energy for the confined fluid is the 
free energy of the uniform fluid of the same density, which means
\begin{eqnarray}
 \label{eq:FinLDA}
 F_{HKM}[\rho]\approx\int d^3r\,\rho(\vec{r})f_{exp}[\rho(\vec{r})].
\end{eqnarray} 
This is the local density approximation 
(LDA), well-known in electronic DFT, but in our case for the Hohenberg-Kohn and Mermin free energy functional. 
It should be noted that the LDA functional goes beyond the CLDFT strategy because it 
is exact in the uniform fluid limit.
To go beyond the LDA, we employ the weighted local density approximation (WLDA) 
\begin{eqnarray}
 \label{eq:FinWLDA}
 F_{HKM}[\rho]&&\approx-\frac{1}{\beta}\int d^3r\,\rho(\vec{r})\log\left(\frac{Z_{kin}}{\rho(\vec{r})V}\right)
 +F_{H}[\rho]+F_{xc}^{LIE-1}[\rho,\overline{\rho}].
\end{eqnarray} 
The first two terms in equation \ref{eq:FinWLDA} were chosen in order
to fulfil the uniform fluid limit exactly. We will call this functional cLIE1 functional 
(classical LIE1), because we will see that it only reflects the classical LIE1 limit.
The approximations \ref{eq:FinLDA}, \ref{eq:FinWLDA} and \ref{eq:qLIE1_1} (in connection with equation \ref{eq:qLIE1_2}) 
were implemented into a new software that allows the minimization of the grand potential. The software is published at our website 
http://www.jacobs-university.de/ses/theine/research. Other functional choices are definitely possible and we 
plan to evaluate more sophisticated approximations than LDA, cLIE1 or qLIE1.      
\subsection{Grand canonical ensemble and external pressure}
The choice of the ensemble should not influence the result in the 
thermodynamic limit. From that point of view it does not matter 
if the method is formulated in the canonical or grand canonical 
ensemble. But the standard methods like GCMC or PI-GCMC\cite{PIGCMC1997} typically use 
the grand canonical ensemble, and thus this choice is more convenient to directly compare the 
QLDFT results with other theoretical approaches.
In the grand canonical ensemble the external pressure and the chemical 
potential can be related to each other directly. This can be achieved most conveniently 
by using the available data from experiments for the uniform fluid. 

It is straight-forward to obtain the free energy of the uniform fluid 
from experiment over a wide range of pressures and 
temperatures \cite{QLDFT2009}, and also the molar volume is available. This 
information is sufficient to relate the chemical potential, pressure and 
temperature using standard thermodynamic relations.

For a given input pressure and temperature we can calculate the free energy $F_N$
per $N$ particles and the molar volume $V_N$ from experiment.
Thus, we obtain the grand potential $\Omega_N$ per mole from  
$\Omega_N = -pV_N$. The chemical potential is given by 
\begin{eqnarray}
\mu =\frac{F_N -\Omega_N}{N}
\end{eqnarray}
\subsection{Minimization within the Car-Parrinello scheme}    
For the minimization of the grand potential $\Omega$ we have chosen the 
Car-Parrinello (CP) scheme \cite{Car1985}. This scheme was originally introduced 
to allow molecular dynamics simulations based on first principles forces on the atoms. However, this scheme can be used to minimize any functional. 
The CP method is often 
discussed in the literature (for an overview about the basic techniques 
see for example ref. \cite{tuckerman1994}), and we only show the main equations 
here.
First, we discuss the minimization of an approximation of the grand potential 
$\Omega$. A grid is covering the space in the simulation box. We denote a grid point with $R$, for the corresponding density at this 
point we use $\rho_{R}$ and for the wave function $\phi_{R}$. In case of the grand canonical ensemble 
we do not have to fix the number of particles, but we have to take care that 
the density at a grid point $R$ cannot be negative. This constraint is already 
fulfilled because $\rho_{R}$ is given by the square of the non-normalized 
wave function $\phi_{R}$, which means $\rho_{R}:=\phi_{R}^2$.

Similar to Car and Parrinello we introduce a fictitious kinetic energy term 
for $\phi_{R}$. The Lagrange function is given by
\begin{eqnarray}
 \mathcal{L}=\frac{m_{x}}{2}\sum_{R}\,\dot{\phi}_{R}^{2}-\Omega[{\phi_{R}}]
\end{eqnarray}    
where $m_{x}$ is the fictitious mass parameter. 
Using the Euler-Lagrange equations we obtain
\begin{eqnarray}
  m_{x}\ddot{\phi}_{R}=-\frac{\partial\Omega}{\partial \phi_{R}}-m_{x}\alpha\dot{x}_{R}.
\end{eqnarray}
Here we have added a friction term $-m_{x}\alpha\dot{\phi}_{R}$ with 
friction parameter $\alpha$. 
The set of the differential equations can  
be discretized using the Verlet algorithm \cite{Verlet1967,Verlet1968}, where the time derivatives $\dot \phi_{R}$ and $\ddot \phi_{R}$ are replaced by  
\begin{equation}
  \dot \phi_{R}\approx\frac{\phi_{R}(+)-\phi_{R}(-)}{2\Delta}
\end{equation}
and 
\begin{equation}
 \ddot \phi_{R}\approx\frac{\phi_{R}(+)-2\phi_{R}(0)+\phi_{R}(-)}{\Delta^2},
\end{equation}
respectively. $(0)$ denotes the present, $(+)$ the next and $(-)$ the previous iteration, differing by time step $\Delta$. 
We can easily rearrange the terms to obtain $\phi_{R}$ for the next step 
\begin{eqnarray}
 \phi_{R}(+)&&=
\left(1+\frac{m_x \alpha}{2 \Delta}\right)^{-1}  
 \nonumber\\
 &&\times\left[-\frac{\Delta^2}{m}\frac{\partial\Omega}{\partial \phi_{R}}+2\phi_{R}(0)-(1-\frac{m_{x}\alpha}{2\Delta})\phi_{R}(-)\right].
\end{eqnarray}  
Note that an explicit calculation of constraints is not necessary in the suggested 
formalism. Choosing a reasonable time step $\Delta$ and friction parameter $\alpha$, the scheme 
yields the minimum of the functional $\Omega$. From now on we will call this 
treatment the CP-GC-QLDFT method.
\section{Benchmark Applications}
To test the presented method we applied it to the adsorption of molecular hydrogen in three selected well known systems that have been 
studied intensively in the literature. The first one is the hard-wall slit pore model potential, the second density fluctuations around a fixed hydrogen probe potential and the third one the graphite slit pore with variable pore size. Further applications 
employing CP-GC-QLDFT to MOFs are currently in progress.
\subsection{Model systems}
The hard-wall slit pore has been widely studied 
in the literature (see for example \cite{hansen2006theory}) and can safely be regarded as the benchmark system for the adsorption of molecular hydrogen.   
The particle-in-a-box character of the boundary implies strong fluctuations of the hydrogen density and thus marks a challenge for the 
numerical approach. The strong oscillations of the density cannot be described well within the LDA. However, satisfactory results 
can be obtained employing the WLDA, as already shown earlier.\cite{QLDFT2009} 
We use these systems to test our CP-GC-QLDFT method thoroughly: We benchmark the results by comparing the hydrogen density profiles to those 
obtained with the C-QLDFT method, and we evaluate the computational performance in terms of convergence and computer time. 

The central potential describes 
a fixed hydrogen molecule at the center of the unit cell. This model allows us to study the density oscillations of the quantum liquid associated to the \lq first solvation shell\rq\ around a 
fixed fictitious hydrogen molecule.  

\begin{figure}[h!]
\begin{center}
\includegraphics[scale=0.85,clip]{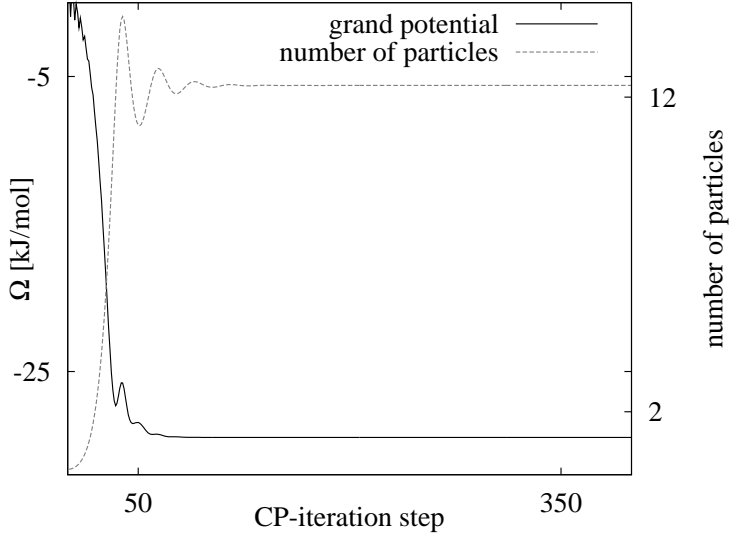}
\caption{\label{fig:minimization} Grand potential $\Omega$ (left $y$-axis) in 
dependence of the simulation step. The grand potential and the number 
of particles (right $y$-axis) show convergence after 400 steps (the time step was 
chosen to be $\Delta=0.05$) for the presented example. The calculation refers to a 
hard-wall slit pore at $T=$100~K and $p$=1 kbar.}
\end{center}
\end{figure}

The convergence of the CP-GC-QLDFT method is illustrated in Figure \ref{fig:minimization} for an exemplary slit pore.
We have used a temperature of 100~K and a pressure of 1~kbar, and employed the cLIE1 functional. The size of the unit cell was chosen 
to be $a=b=5$\AA and $c=20$\AA. A grid of 1 point per \AA in $x$ and $y$ direction and 
10 points per \AA in $z$ direction was applied. As shown in Figure \ref{fig:minimization}, after approximately 
400 simulation steps (we use $\Delta=0.05$ for the presented example) the change in the grand potential becomes negligible. 
After $\sim$ 200 steps the number of particles in the unit cell changes by $\Delta N=10^{-2}$ particles, after $\sim$ 300 
steps by $\Delta N=10^{-4}$. We consider the system to be converged by $\Delta N=10^{-5}$ which is obtained after $\sim$ 400 
steps.

It is not straight-forward to compare the grand canonical and the canonical ensemble: While in C-QLDFT the number of particles in the 
simulation box is fixed, in GC-QLDFT we fix the external pressure or, equivalently, the chemical potential. In order to achieve 
the best possible comparison, we have first converged the system in the CP-GC-QLDFT approach, and then used the obtained molar volume 
as input variable for C-QLDFT. Calculations have been carried out for $T$=100~K, 200~K and 300~K. In case of CP-GC-QLDFT we have applied the 
LDA, cLIE1 and qLIE1 functionals in case of C-QLDFT the LIE0 (for details see \cite{QLDFT2009}) and LIE1 
functionals. Calculations have been carried out for the same setup and temperatures as reported above.

\begin{figure*}[h!]
\begin{center}
\includegraphics[scale=1.0,clip]{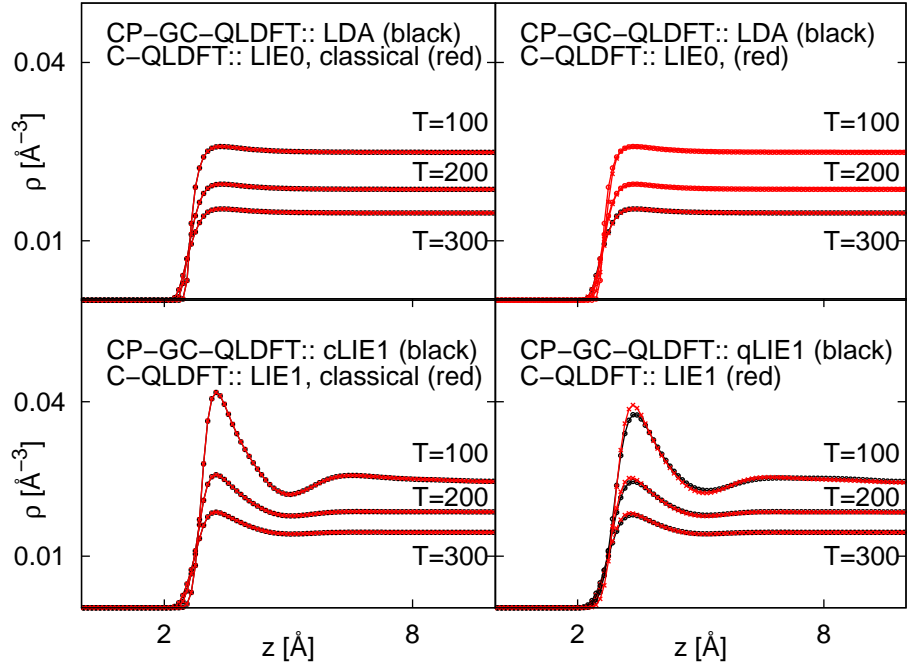}
\caption{\label{fig:densitycentral} Density profiles of the hydrogen quantum liquid in the central potential obtained with the 
C-QLDFT and CP-GC-QLDFT methods. A pressure of 1 kbar was chosen for the CP-GC-QLDFT calculations. On the left-hand side 
it is shown that the cLIE1 (LDA) density profile is nearly the same compared to the classical C-QLDFT LIE1 
(LIE0) result. On the right-hand panel we compare the LDA and qLIE1 density profiles with the results obtained 
from non-classical LIE0 and LIE1 calculations.}
\end{center}
\end{figure*}

In Figure \ref{fig:densitycentral} the density profiles of both methods are compared for the central potential. 
As expected, within the LDA we obtain a nearly shapeless density profile. The LDA, where the $v_{12}$ 
intermolecular potential is incorporated in the HKM operator, and LIE0, that is LDA with neglection of the interparticle interactions of the fluid, give - not surprisingly -  
very similar results. 

The non-physical absence of the density oscillations has already been discussed in detail earlier.\cite{QLDFT2009} 
Here, we note that it is irrelevant if the local approximation is made to the classical or the quantum-mechanical system and 
state that it should only be applied if the potential variation and the associated density oscillations are 
expected to be very small. 

Because of the fact that the magnitude of the density oscillations is controlled by the choice of the 
weighting function $w$ used to construct $\overline{\rho}$, we are not surprised that the densities of the 
cLIE1, qLIE1 and the LIE-1 functionals are in rather close agreement with each other. 

\begin{figure*}[h!]
\begin{center}
\includegraphics[scale=1.0,clip]{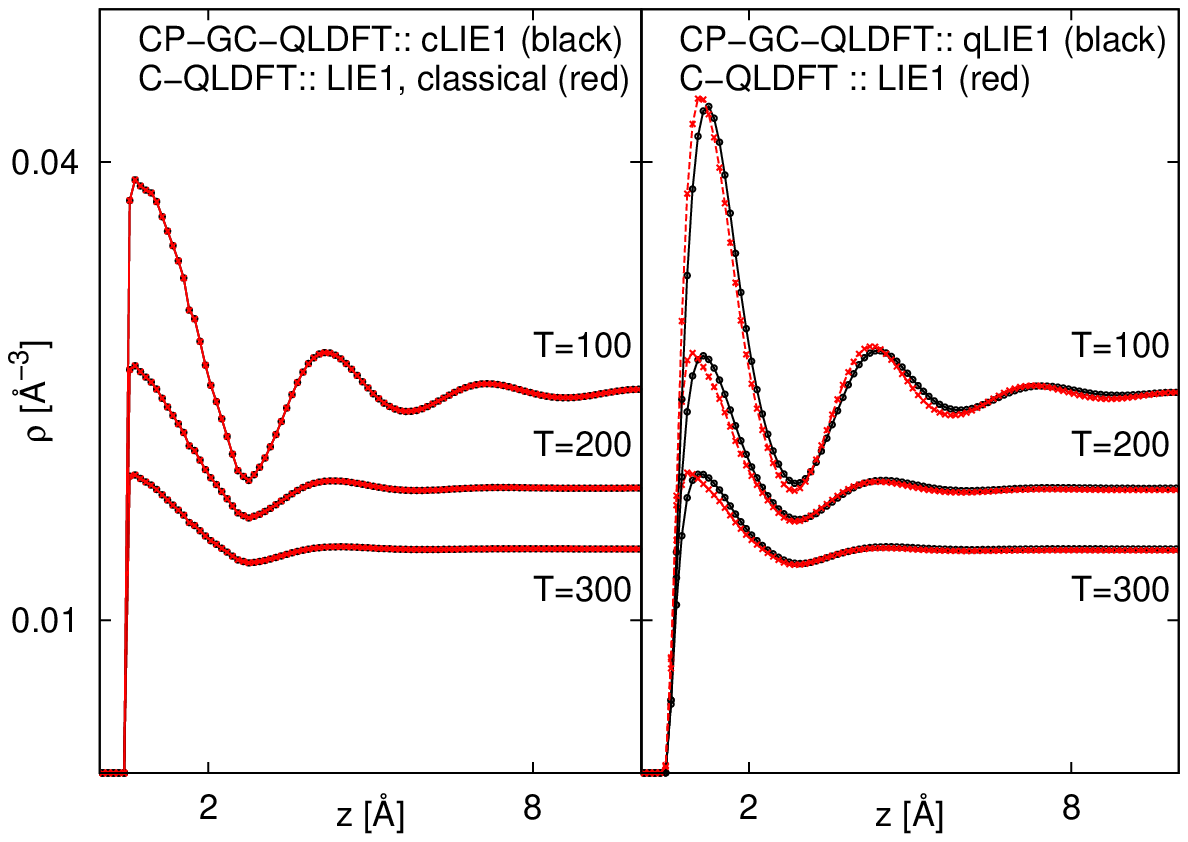}
\caption{\label{fig:densityhardwall} Density profiles of the hydrogen quantum liquid for the hard-wall potential. The position of the hard wall 
is $z=0.5$\,\AA\, for the presented example. The CP-GC-QLDFT result ($p$=1 kbar) obtained with the cLIE1 functional 
(left-hand side) shows a discontinuity at the hard wall in contrast to the result obtained with qLIE1 (right-hand side). The LDA result for 
the hard-wall potential is not presented. In that case one obtains a step-like density profile without a 
quantum correction close to the hard wall.}    
\end{center}
\end{figure*}

The classical LIE1 functional produces a discontinuity of the density profile for the hard wall potential 
(see figure \ref{fig:densityhardwall}). This behaviour is due to the discontinuous character of the model potential and
 thus will not occur when treating real systems with parametrized force fields describing the host-guest interaction.

In contrast to the classical functional (cLIE1), the qLIE1 correctly describes the density oscillations and also the quantum 
behaviour in a satisfactory manner: in comparison to the C-QLDFT result, small differences in the density profile are 
observed close to the hard walls.
We observe a slight overdelocalization in qLIE1 due to the choice of the zero temperature reference system, a consequence of 
the missing higher-energy one-particle states.
In contrast to the cLIE1 approximation, the qLIE1 functional is able to describe the density oscillations and also the quantum 
behaviour in a satisfactory manner. In comparison to the C-QLDFT result, small differences in the density profile can be 
observed close to the hard walls.
The zero temperature reference system delocalizes 
the particles a bit too much which is a consequence of the missing higher-energy one-particle states. The differences have, 
however, only numerical character and if necessary one can change the weighting function used in the qLIE1 functional 
to obtain a better agreement. In future we plan to analyse this behaviour in more detail by comparing our results with PI calculations.             

At this point we compare the computational performance of C-QLDFT with CP-GC-QLDFT. A big advantage of the CP-GC-QLDFT treatment 
is that the method is formulated in a way that the speed of the calculations does not depend strongly on the temperature. In the 
exponential series employed to calculate the partition function in C-QLDFT\cite{patchkovskii2007}, low temperatures 
result in non-sparse Hamiltonians and long series are necessary to reach convergence. In our direct 
minimization scheme there is no explicit dependency on the temperature, and hence this scheme works with 
about the same performance for all temperatures tested. It should be noted that for a calculation with $T$=50 K 
the computer time on the same machine accounts for a few seconds for the CP-GC-QLDFT scheme, compared to more 
than a week using the C-QLDFT scheme, employing the exponential series. For high temperatures (e.g. 300~K) both codes show comparable 
performance. 
\subsection{Carbon layers}         
Experiments in the late 1990s have reported that porous carbon structures 
can store a significant amount of hydrogen even at room temperature 
\cite{dillon1997,Chambers1998}. These experiments have motivated a lot of 
theoretical and experimental investigations of carbon based materials in 
the last decades. Today it is known that these experiments were not reproducible 
and the hydrogen uptake of these materials is only significant at much lower 
temperatures. However, carbon based materials are still among the attractive candidates for 
hydrogen storage. For reviews about that topic we refer to \cite{hirscher2009,Parrinello2001}.

For our benchmark applications we will focus on graphite slit pores (GSPs). 
GSPs have been studied theoretically with GCMC calculations by Aga et al. \cite{Aga2007} 
and Rzepka et al. \cite{rzepka1998}, with the PI-GCMC method \cite{PIGCMC1997} by Wang 
and Johnson \cite{johnsonh2carbonfoam} (idealized carbon slit pore) and also with 
a free energy based method by Patchkovskii et al. \cite{PNAS2005}. 
Today it is well known that a change of the graphite interlayer distance influences the 
hydrogen uptake significantly, depending on external pressure and temperature.

For the theoretical treatment of GSPs, different $C-H_2$ potentials have been employed 
in the past. Wang et al. \cite{Wang1980} suggested a Lennard-Jones potential (LJ) 
\begin{eqnarray}
 V(r)=4\varepsilon\left[\left(\frac{\sigma}{r}\right)^{12}-
 \left(\frac{\sigma}{r}\right)^{6}\right]
\end{eqnarray}
with $\sigma$ = 2.97 \AA\ and $\varepsilon$ = 3.69~meV, while Patchkovskii 
et al. \cite{PNAS2005} have chosen a Buckingham potential (B)                      
\begin{eqnarray}
 V(r)=Ae^{-\alpha r}+C_{6}r^{-6}
\end{eqnarray}
with $A$=1099.52~eV, $\alpha$ = 3.5763 \AA$^{-1}$ and $C_{6}$ = -17.36~eV \AA$^6$. It was 
already discussed that the B potential predicts a significantly stronger $C-H_2$ 
binding energy than the LJ potential with the parameters reported above, and therefore predicts a larger amount of hydrogen uptake.\cite{Aga2007}   

It should be mentioned that the PI-GCMC calculations of Wang and Johnson 
\cite{johnsonh2carbonfoam} were performed with the Crowell-Brown (CB) potential \cite{crowell1982}. 
This potential depends on the number and distance of graphene planes. As we wanted to apply potentials 
which can be used in a more general way for further applications, we decided to 
focus on the LJ and B potentials only. 
\begin{figure*}
\begin{center}
\includegraphics[scale=1.0,clip]{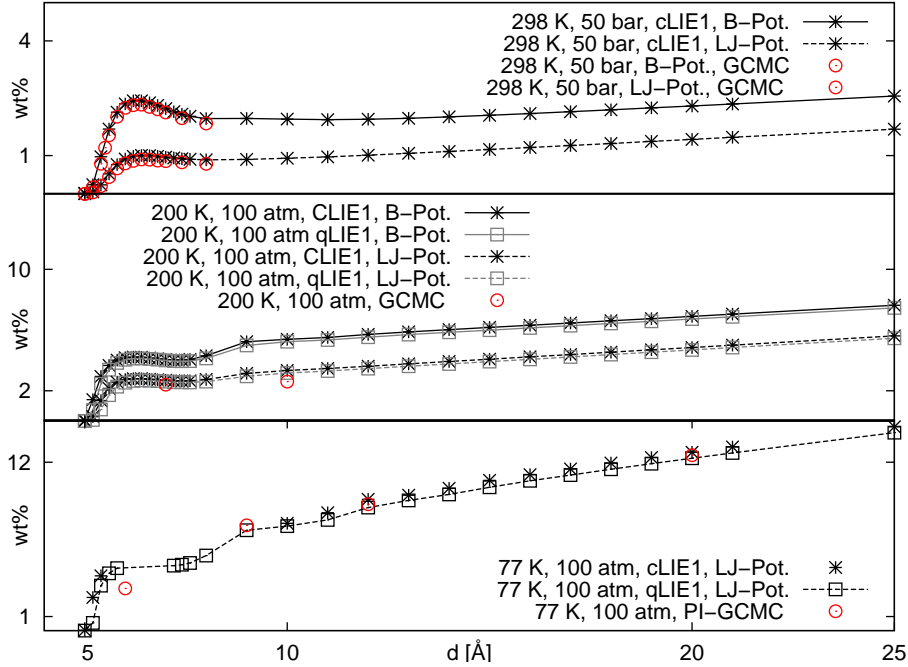}
\caption{\label{fig:wpercent} Weight percent adsorption in dependence of the interlayer distance $d$. 
Results for $T$=298 K at $p$=50 bar, as well as for $T$=77 K and 200 K at $p$=101.325 bar are shown. The results are compared with 
GCMC calculations of Aga and co-workers \cite{Aga2007} ($T$=300 K), with the GCMC results of Rzepka et al. 
\cite{rzepka1998} ($T$=200 K) and with the PI-GCMC calculations of Wang and Johnson \cite{johnsonh2carbonfoam}.}
\end{center}
\end{figure*}
The size of the GSP unit cell 
was chosen to be $a$ = 12.30622099 \AA, $b$ = 12.789 \AA and $c$ = 4$d$. $d$ denotes the 
distance between the graphene layers. We used 6 grid points per \AA for our calculations.
The gravimetric adsorption capacity (in weight percent wt \%) is shown in Figure \ref{fig:wpercent}. Simulations have been carried out 
at $T$=77~K, 200~K and 298~K. In case of 77~K and 200~K we applied $p$= 101.325 bar, in case of 
$T$=298~K $p$=50 bar. In all cases, the chosen force field has an enormous influence on the 
predicted hydrogen storage capacity. It illustrates the importance of high precision host-guest interaction potentials. 
Our CP-GC-QLDFT results with cLIE1 functional match Aga et al.'s GCMC calculations \cite{Aga2007} very closely at 
room temperature. We observe the same maximum between 6.2 and 6.6 \AA, caused by the superposition of 
the host-guest interaction of upper and lower layer. Like GCMC, CP-GC-QLDFT also predicts a vanishing hydrogen density 
at approximatly 5 \AA. 
       
The qLIE1 functional predicts a smaller hydrogen uptake for both potentials. This can be expected because 
quantum effects lower the heat of adsorption. Our results match the GCMC calculations of Rzepka et al. \cite{rzepka1998} 
and the PI-GCMC calculations of Wang and Johnson \cite{johnsonh2carbonfoam} closely. We assume that different 
potentials and the rather different numerical implementation explain the deviations.

For the low-temperature and high pressure simulation we have applied the LIE1 functional in the classical (cLIE1) and quantum-mechanical (qLIE1) variants. 
In case of $T$=77~K we focus on the LJ-potential. We have also done calculations with the B-potential, but our 
simulations yield results outside the safety range of the available experimental data that is used to determine $F_{xc}$. This is not a 
problem of the presented method, and should 
be solved by implementing more accurate experimental data.

\begin{figure}[h!]
\begin{center}
\includegraphics[scale=0.2,clip]{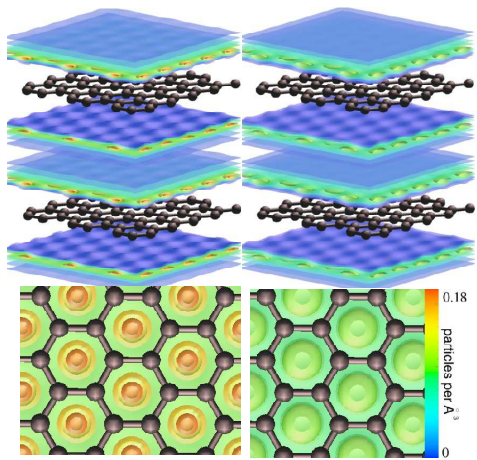}
\caption{\label{fig:densgraph} Density profiles for $T$=77~K and $p$= 101.325 bar obtained from CP-GC-QLDFT 
calculations using the LJ potential. The distance of the graphene layers is 10 \AA. 
On the left-hand 
side the cLIE1 result is shown, on the right side the qLIE1 result.}
\end{center}
\end{figure}

The density profiles obtained with the cLIE1 and qLIE1 functionals at $T$=77~K and $p$= 101.325 bar 
are shown in figure \ref{fig:densgraph}. For both functionals we find spherical (disc-like) areas where a 
large amount of hydrogen is predicted. These high-density areas are a consequence of the external potential which shows the same pattern, as already observed earlier.\cite{PNAS2005}        
\section{Conclusions}
We have reformulated Quantized Liquid Density-Functional Theory in the grand canonical ensemble. The method has been implemented
employing a Car-Parrinello scheme for direct minimization of the grand potential. We obtain a significant computational performance 
for low-temperature applications. We have compared our implementation with the canonical version of QLDFT for model slit pores and 
with GCMC calculations for graphite slit pores. In future work we plan to examine more sophisticated approximations of the 
free energy functional. The software used for this work is available on our website at http://www.jacobs-university.de/ses/theine/research.    
\section{Acknowledgements}
Financial support by Deutsche Forschungsgemeinschaft (HE 3543/7-2) within Priority Program SPP1613 is gratefully acknowledged. 
We thank Prof. Charusita Chakravarty for inspiring discussions.  
\providecommand*{\mcitethebibliography}{\thebibliography}
\csname @ifundefined\endcsname{endmcitethebibliography}
{\let\endmcitethebibliography\endthebibliography}{}


\begin{mcitethebibliography}{45}
\providecommand*{\natexlab}[1]{#1}
\providecommand*{\mciteSetBstSublistMode}[1]{}
\providecommand*{\mciteSetBstMaxWidthForm}[2]{}
\providecommand*{\mciteBstWouldAddEndPuncttrue}
  {\def\EndOfBibitem{\unskip.}}
\providecommand*{\mciteBstWouldAddEndPunctfalse}
  {\let\EndOfBibitem\relax}
\providecommand*{\mciteSetBstMidEndSepPunct}[3]{}
\providecommand*{\mciteSetBstSublistLabelBeginEnd}[3]{}
\providecommand*{\EndOfBibitem}{}
\mciteSetBstSublistMode{f}
\mciteSetBstMaxWidthForm{subitem}{(\alph{mcitesubitemcount})}
\mciteSetBstSublistLabelBeginEnd{\mcitemaxwidthsubitemform\space}
{\relax}{\relax}

\bibitem[Rosi et~al.(2003)Rosi, Eckert, Eddaoudi, Vodak, Kim, O'Keeffe, and
  Yaghi]{h2mof}
Rosi,~N.; Eckert,~J.; Eddaoudi,~M.; Vodak,~D.; Kim,~J.; O'Keeffe,~M.; Yaghi,~O.
  \emph{Science} \textbf{2003}, \emph{300}, 1127--1129\relax
\mciteBstWouldAddEndPuncttrue
\mciteSetBstMidEndSepPunct{\mcitedefaultmidpunct}
{\mcitedefaultendpunct}{\mcitedefaultseppunct}\relax
\EndOfBibitem
\bibitem[Chen et~al.(2008)Chen, Zhao, Putkham, Hong, Lobkovsky, Hurtado,
  Fletcher, and Thomas]{h2d2-japanese-jacs}
Chen,~B.; Zhao,~X.; Putkham,~A.; Hong,~K.; Lobkovsky,~E.~B.; Hurtado,~E.~J.;
  Fletcher,~A.~J.; Thomas,~K.~M. \emph{Journal of the American Chemical
  Society} \textbf{2008}, \emph{130}, 6411--6423, PMID: 18435535\relax
\mciteBstWouldAddEndPuncttrue
\mciteSetBstMidEndSepPunct{\mcitedefaultmidpunct}
{\mcitedefaultendpunct}{\mcitedefaultseppunct}\relax
\EndOfBibitem
\bibitem[Teufel et~al.(2013)Teufel, Oh, Hirscher, Wahiduzzaman, Zhechkov, Kuc,
  Heine, Denysenko, and Volkmer]{Teufel2013}
Teufel,~J.; Oh,~H.; Hirscher,~M.; Wahiduzzaman,~M.; Zhechkov,~L.; Kuc,~A.;
  Heine,~T.; Denysenko,~D.; Volkmer,~D. \emph{Advanced Materials}
  \textbf{2013}, \emph{25}, 635--639\relax
\mciteBstWouldAddEndPuncttrue
\mciteSetBstMidEndSepPunct{\mcitedefaultmidpunct}
{\mcitedefaultendpunct}{\mcitedefaultseppunct}\relax
\EndOfBibitem
\bibitem[Cai et~al.(2012)Cai, Xing, and Zhao]{cai2012}
Cai,~J.; Xing,~Y.; Zhao,~X. \emph{RSC Advances} \textbf{2012}, \emph{2},
  8579--8586\relax
\mciteBstWouldAddEndPuncttrue
\mciteSetBstMidEndSepPunct{\mcitedefaultmidpunct}
{\mcitedefaultendpunct}{\mcitedefaultseppunct}\relax
\EndOfBibitem
\bibitem[Liu et~al.(2012)Liu, Wang, Mi, Zhong, Yang, and D]{liu2012}
Liu,~D.; Wang,~W.; Mi,~J.; Zhong,~C.; Yang,~Q.; D,~W. \emph{Industrial and
  Engineering Chemistry Research} \textbf{2012}, \emph{51}, 434--442\relax
\mciteBstWouldAddEndPuncttrue
\mciteSetBstMidEndSepPunct{\mcitedefaultmidpunct}
{\mcitedefaultendpunct}{\mcitedefaultseppunct}\relax
\EndOfBibitem
\bibitem[Beenakker et~al.(1995)Beenakker, Borman, and Krylov]{Beenakker1995}
Beenakker,~J. J.~M.; Borman,~V.~D.; Krylov,~S.~Y. \emph{Chem. Phys. Lett.}
  \textbf{1995}, \emph{232}, 379--382\relax
\mciteBstWouldAddEndPuncttrue
\mciteSetBstMidEndSepPunct{\mcitedefaultmidpunct}
{\mcitedefaultendpunct}{\mcitedefaultseppunct}\relax
\EndOfBibitem
\bibitem[Garberoglio et~al.(2005)Garberoglio, Skoulidas, and
  Johnson]{Garberoglio2005}
Garberoglio,~G.; Skoulidas,~A.~I.; Johnson,~J.~K. \emph{The Journal of Physical
  Chemistry B} \textbf{2005}, \emph{109}, 13094--13103, PMID: 16852629\relax
\mciteBstWouldAddEndPuncttrue
\mciteSetBstMidEndSepPunct{\mcitedefaultmidpunct}
{\mcitedefaultendpunct}{\mcitedefaultseppunct}\relax
\EndOfBibitem
\bibitem[Kumar et~al.(2006)Kumar, Jobic, and Bhatia]{Kumar2006}
Kumar,~A. V.~A.; Jobic,~H.; Bhatia,~S.~K. \emph{The Journal of Physical
  Chemistry B} \textbf{2006}, \emph{110}, 16666--16671, PMID: 16913804\relax
\mciteBstWouldAddEndPuncttrue
\mciteSetBstMidEndSepPunct{\mcitedefaultmidpunct}
{\mcitedefaultendpunct}{\mcitedefaultseppunct}\relax
\EndOfBibitem
\bibitem[Feynman and Hibbs(1965)]{FeynmanHibbs}
Feynman,~R.~P.; Hibbs,~A.~R. \emph{Quantum Mechanics and Path Integrals};
\newblock New York: McGraw-Hill, 1965\relax
\mciteBstWouldAddEndPuncttrue
\mciteSetBstMidEndSepPunct{\mcitedefaultmidpunct}
{\mcitedefaultendpunct}{\mcitedefaultseppunct}\relax
\EndOfBibitem
\bibitem[Feynman(1975)]{FeynmanSM}
Feynman,~R.~P. \emph{Statistical Mechanics};
\newblock Benjamin: New-York, 1975\relax
\mciteBstWouldAddEndPuncttrue
\mciteSetBstMidEndSepPunct{\mcitedefaultmidpunct}
{\mcitedefaultendpunct}{\mcitedefaultseppunct}\relax
\EndOfBibitem
\bibitem[Feynman(1948)]{PI1948}
Feynman,~R.~P. \emph{Rev. Mod. Phys.} \textbf{1948}, \emph{20}, 367--387\relax
\mciteBstWouldAddEndPuncttrue
\mciteSetBstMidEndSepPunct{\mcitedefaultmidpunct}
{\mcitedefaultendpunct}{\mcitedefaultseppunct}\relax
\EndOfBibitem
\bibitem[Chakravarty(1995)]{chakravarty1995}
Chakravarty,~C. \emph{Molecular Physics} \textbf{1995}, \emph{84},
  845--852\relax
\mciteBstWouldAddEndPuncttrue
\mciteSetBstMidEndSepPunct{\mcitedefaultmidpunct}
{\mcitedefaultendpunct}{\mcitedefaultseppunct}\relax
\EndOfBibitem
\bibitem[Wang and Johnson(1999)]{johnsonh2carbonfoam}
Wang,~Q.; Johnson,~J.~K. \emph{The Journal of Chemical Physics} \textbf{1999},
  \emph{110}, 577--586\relax
\mciteBstWouldAddEndPuncttrue
\mciteSetBstMidEndSepPunct{\mcitedefaultmidpunct}
{\mcitedefaultendpunct}{\mcitedefaultseppunct}\relax
\EndOfBibitem
\bibitem[Ceperley(1995)]{ceperley1995}
Ceperley,~D. \emph{Reviews of Modern Physics} \textbf{1995}, \emph{67},
  279\relax
\mciteBstWouldAddEndPuncttrue
\mciteSetBstMidEndSepPunct{\mcitedefaultmidpunct}
{\mcitedefaultendpunct}{\mcitedefaultseppunct}\relax
\EndOfBibitem
\bibitem[Chakravarty(1997)]{chakravarty1997}
Chakravarty,~C. \emph{International reviews in physical chemistry}
  \textbf{1997}, \emph{16}, 421--444\relax
\mciteBstWouldAddEndPuncttrue
\mciteSetBstMidEndSepPunct{\mcitedefaultmidpunct}
{\mcitedefaultendpunct}{\mcitedefaultseppunct}\relax
\EndOfBibitem
\bibitem[Patchkovskii et~al.(2005)Patchkovskii, Tse, Yurchenko, Zhechkov,
  Heine, and Seifert]{PNAS2005}
Patchkovskii,~S.; Tse,~J.~S.; Yurchenko,~S.~N.; Zhechkov,~L.; Heine,~T.;
  Seifert,~G. \emph{Proceedings of the National Academy of Sciences of the
  United States of America} \textbf{2005}, \emph{102}, 10439--10444\relax
\mciteBstWouldAddEndPuncttrue
\mciteSetBstMidEndSepPunct{\mcitedefaultmidpunct}
{\mcitedefaultendpunct}{\mcitedefaultseppunct}\relax
\EndOfBibitem
\bibitem[Patchkovskii and Heine(2007)]{patchkovskii2007}
Patchkovskii,~S.; Heine,~T. \emph{Phys. Chem. Chem. Phys.} \textbf{2007},
  \emph{9}, 2697--2705\relax
\mciteBstWouldAddEndPuncttrue
\mciteSetBstMidEndSepPunct{\mcitedefaultmidpunct}
{\mcitedefaultendpunct}{\mcitedefaultseppunct}\relax
\EndOfBibitem
\bibitem[Kohn and Sham(1965)]{KS1965}
Kohn,~W.; Sham,~L. \emph{Physical Review} \textbf{1965}, \emph{140}, 1--5\relax
\mciteBstWouldAddEndPuncttrue
\mciteSetBstMidEndSepPunct{\mcitedefaultmidpunct}
{\mcitedefaultendpunct}{\mcitedefaultseppunct}\relax
\EndOfBibitem
\bibitem[Patchkovskii and Heine(2009)]{QLDFT2009}
Patchkovskii,~S.; Heine,~T. \emph{Phys. Rev. E} \textbf{2009}, \emph{80},
  031603\relax
\mciteBstWouldAddEndPuncttrue
\mciteSetBstMidEndSepPunct{\mcitedefaultmidpunct}
{\mcitedefaultendpunct}{\mcitedefaultseppunct}\relax
\EndOfBibitem
\bibitem[Martinez-Mesa et~al.(2011)Martinez-Mesa, Yurchenko, Patchkovskii,
  Heine, and Seifert]{Mesa2011}
Martinez-Mesa,~A.; Yurchenko,~S.~N.; Patchkovskii,~S.; Heine,~T.; Seifert,~G.
  \emph{The Journal of Chemical Physics} \textbf{2011}, \emph{135},
  214701\relax
\mciteBstWouldAddEndPuncttrue
\mciteSetBstMidEndSepPunct{\mcitedefaultmidpunct}
{\mcitedefaultendpunct}{\mcitedefaultseppunct}\relax
\EndOfBibitem
\bibitem[Martinez-Mesa et~al.(2012)Martinez-Mesa, Zhechkov, Yurchenko, Heine,
  Seifert, and Rubayo-Soneira]{Mesa2012}
Martinez-Mesa,~A.; Zhechkov,~L.; Yurchenko,~S.~N.; Heine,~T.; Seifert,~G.;
  Rubayo-Soneira,~J. \emph{The Journal of Physical Chemistry C} \textbf{2012},
  \emph{116}, 19543--19553\relax
\mciteBstWouldAddEndPuncttrue
\mciteSetBstMidEndSepPunct{\mcitedefaultmidpunct}
{\mcitedefaultendpunct}{\mcitedefaultseppunct}\relax
\EndOfBibitem
\bibitem[Lukose et~al.(2012)Lukose, Wahiduzzaman, Kuc, and Heine]{Lukose2012}
Lukose,~B.; Wahiduzzaman,~M.; Kuc,~A.; Heine,~T. \emph{The Journal of Physical
  Chemistry C} \textbf{2012}, \emph{116}, 22878--22884\relax
\mciteBstWouldAddEndPuncttrue
\mciteSetBstMidEndSepPunct{\mcitedefaultmidpunct}
{\mcitedefaultendpunct}{\mcitedefaultseppunct}\relax
\EndOfBibitem
\bibitem[Car and Parrinello(1985)]{Car1985}
Car,~R.; Parrinello,~M. \emph{Phys. Rev. Lett.} \textbf{1985}, \emph{55},
  2471--2474\relax
\mciteBstWouldAddEndPuncttrue
\mciteSetBstMidEndSepPunct{\mcitedefaultmidpunct}
{\mcitedefaultendpunct}{\mcitedefaultseppunct}\relax
\EndOfBibitem
\bibitem[Hohenberg and Kohn(1964)]{Hohenberg1964}
Hohenberg,~P.; Kohn,~W. \emph{Phys. Rev.} \textbf{1964}, \emph{136},
  B864--B871\relax
\mciteBstWouldAddEndPuncttrue
\mciteSetBstMidEndSepPunct{\mcitedefaultmidpunct}
{\mcitedefaultendpunct}{\mcitedefaultseppunct}\relax
\EndOfBibitem
\bibitem[Mermin(1965)]{Mermin1965}
Mermin,~N.~D. \emph{Phys. Rev.} \textbf{1965}, \emph{137}, A1441--A1443\relax
\mciteBstWouldAddEndPuncttrue
\mciteSetBstMidEndSepPunct{\mcitedefaultmidpunct}
{\mcitedefaultendpunct}{\mcitedefaultseppunct}\relax
\EndOfBibitem
\bibitem[Diep and Johnson(2000)]{diep2000}
Diep,~P.; Johnson,~J.~K. \emph{The Journal of Chemical Physics} \textbf{2000},
  \emph{112}, 4465\relax
\mciteBstWouldAddEndPuncttrue
\mciteSetBstMidEndSepPunct{\mcitedefaultmidpunct}
{\mcitedefaultendpunct}{\mcitedefaultseppunct}\relax
\EndOfBibitem
\bibitem[Curtin and Ashcroft(1985)]{WLDA1985}
Curtin,~W.~A.; Ashcroft,~N.~W. \emph{Phys. Rev. A} \textbf{1985}, \emph{32},
  2909--2919\relax
\mciteBstWouldAddEndPuncttrue
\mciteSetBstMidEndSepPunct{\mcitedefaultmidpunct}
{\mcitedefaultendpunct}{\mcitedefaultseppunct}\relax
\EndOfBibitem
\bibitem[McCarty et~al.(1981)McCarty, Hord, and Roder]{mccarty1981}
McCarty,~R.~D.; Hord,~J.; Roder,~H.~M. \emph{Selected properties of hydrogen
  (engineering design data)};
\newblock Technical Report, 1981\relax
\mciteBstWouldAddEndPuncttrue
\mciteSetBstMidEndSepPunct{\mcitedefaultmidpunct}
{\mcitedefaultendpunct}{\mcitedefaultseppunct}\relax
\EndOfBibitem
\bibitem[Mills et~al.(1977)Mills, Liebenberg, Bronson, and Schmidt]{mills1977}
Mills,~R.; Liebenberg,~D.; Bronson,~J.; Schmidt,~L. \emph{The Journal of
  Chemical Physics} \textbf{1977}, \emph{66}, 3076\relax
\mciteBstWouldAddEndPuncttrue
\mciteSetBstMidEndSepPunct{\mcitedefaultmidpunct}
{\mcitedefaultendpunct}{\mcitedefaultseppunct}\relax
\EndOfBibitem
\bibitem[Saam and Ebner(1977)]{saam1977}
Saam,~W.; Ebner,~C. \emph{Physical Review A} \textbf{1977}, \emph{15},
  2566--2568\relax
\mciteBstWouldAddEndPuncttrue
\mciteSetBstMidEndSepPunct{\mcitedefaultmidpunct}
{\mcitedefaultendpunct}{\mcitedefaultseppunct}\relax
\EndOfBibitem
\bibitem[Evans(1979)]{evans1979}
Evans,~R. \emph{Advances in Physics} \textbf{1979}, \emph{28}, 143--200\relax
\mciteBstWouldAddEndPuncttrue
\mciteSetBstMidEndSepPunct{\mcitedefaultmidpunct}
{\mcitedefaultendpunct}{\mcitedefaultseppunct}\relax
\EndOfBibitem
\bibitem[Singh(1991)]{singh1991}
Singh,~Y. \emph{Physics Reports} \textbf{1991}, \emph{207}, 351--444\relax
\mciteBstWouldAddEndPuncttrue
\mciteSetBstMidEndSepPunct{\mcitedefaultmidpunct}
{\mcitedefaultendpunct}{\mcitedefaultseppunct}\relax
\EndOfBibitem
\bibitem[Wang et~al.(1997)Wang, Johnson, and Broughton]{PIGCMC1997}
Wang,~Q.; Johnson,~J.~K.; Broughton,~J.~Q. \emph{The Journal of Chemical
  Physics} \textbf{1997}, \emph{107}, 5108--5117\relax
\mciteBstWouldAddEndPuncttrue
\mciteSetBstMidEndSepPunct{\mcitedefaultmidpunct}
{\mcitedefaultendpunct}{\mcitedefaultseppunct}\relax
\EndOfBibitem
\bibitem[Tuckerman and Parrinello(1994)]{tuckerman1994}
Tuckerman,~M.~E.; Parrinello,~M. \emph{The Journal of Chemical Physics}
  \textbf{1994}, \emph{101}, 1302--1315\relax
\mciteBstWouldAddEndPuncttrue
\mciteSetBstMidEndSepPunct{\mcitedefaultmidpunct}
{\mcitedefaultendpunct}{\mcitedefaultseppunct}\relax
\EndOfBibitem
\bibitem[Verlet(1967)]{Verlet1967}
Verlet,~L. \emph{Phys. Rev.} \textbf{1967}, \emph{159}, 98--103\relax
\mciteBstWouldAddEndPuncttrue
\mciteSetBstMidEndSepPunct{\mcitedefaultmidpunct}
{\mcitedefaultendpunct}{\mcitedefaultseppunct}\relax
\EndOfBibitem
\bibitem[Verlet(1968)]{Verlet1968}
Verlet,~L. \emph{Phys. Rev.} \textbf{1968}, \emph{165}, 201--214\relax
\mciteBstWouldAddEndPuncttrue
\mciteSetBstMidEndSepPunct{\mcitedefaultmidpunct}
{\mcitedefaultendpunct}{\mcitedefaultseppunct}\relax
\EndOfBibitem
\bibitem[Hansen and McDonald(2006)]{hansen2006theory}
Hansen,~J.; McDonald,~I. \emph{Theory of simple liquids};
\newblock Academic press, 2006\relax
\mciteBstWouldAddEndPuncttrue
\mciteSetBstMidEndSepPunct{\mcitedefaultmidpunct}
{\mcitedefaultendpunct}{\mcitedefaultseppunct}\relax
\EndOfBibitem
\bibitem[Dillon et~al.(1997)Dillon, Jones, Bekkedahl, Kiang, Bethune, and
  Heben]{dillon1997}
Dillon,~A.~C.; Jones,~K.~M.; Bekkedahl,~T.~A.; Kiang,~C.~H.; Bethune,~D.~S.;
  Heben,~M.~J. \emph{Nature} \textbf{1997}, \emph{386}, 377--379\relax
\mciteBstWouldAddEndPuncttrue
\mciteSetBstMidEndSepPunct{\mcitedefaultmidpunct}
{\mcitedefaultendpunct}{\mcitedefaultseppunct}\relax
\EndOfBibitem
\bibitem[Chambers et~al.(1998)Chambers, Park, Baker, and
  Rodriguez]{Chambers1998}
Chambers,~A.; Park,~C.; Baker,~R. T.~K.; Rodriguez,~N.~M. \emph{The Journal of
  Physical Chemistry B} \textbf{1998}, \emph{102}, 4253--4256\relax
\mciteBstWouldAddEndPuncttrue
\mciteSetBstMidEndSepPunct{\mcitedefaultmidpunct}
{\mcitedefaultendpunct}{\mcitedefaultseppunct}\relax
\EndOfBibitem
\bibitem[Hirscher et~al.(2009)Hirscher, Z{\"u}ttel, Borgschulte, and
  Schlapbach]{hirscher2009}
Hirscher,~M.; Z{\"u}ttel,~A.; Borgschulte,~A.; Schlapbach,~L. \emph{Ceramic
  Transactions} \textbf{2009}, \emph{202}, year\relax
\mciteBstWouldAddEndPuncttrue
\mciteSetBstMidEndSepPunct{\mcitedefaultmidpunct}
{\mcitedefaultendpunct}{\mcitedefaultseppunct}\relax
\EndOfBibitem
\bibitem[{Meregalli} and {Parrinello}(2001)]{Parrinello2001}
{Meregalli},~V.; {Parrinello},~M. \emph{Applied Physics A: Materials Science \&
  Processing} \textbf{2001}, \emph{72}, 143--146\relax
\mciteBstWouldAddEndPuncttrue
\mciteSetBstMidEndSepPunct{\mcitedefaultmidpunct}
{\mcitedefaultendpunct}{\mcitedefaultseppunct}\relax
\EndOfBibitem
\bibitem[Aga et~al.(2007)Aga, Fu, Kr\ifmmode~\check{c}\else \v{c}\fi{}mar, and
  Morris]{Aga2007}
Aga,~R.~S.; Fu,~C.~L.; Kr\ifmmode~\check{c}\else \v{c}\fi{}mar,~M.;
  Morris,~J.~R. \emph{Phys. Rev. B} \textbf{2007}, \emph{76}, 165404\relax
\mciteBstWouldAddEndPuncttrue
\mciteSetBstMidEndSepPunct{\mcitedefaultmidpunct}
{\mcitedefaultendpunct}{\mcitedefaultseppunct}\relax
\EndOfBibitem
\bibitem[Rzepka et~al.(1998)Rzepka, Lamp, and De~La Casa-Lillo]{rzepka1998}
Rzepka,~M.; Lamp,~P.; De~La Casa-Lillo,~M. \emph{The Journal of Physical
  Chemistry B} \textbf{1998}, \emph{102}, 10894--10898\relax
\mciteBstWouldAddEndPuncttrue
\mciteSetBstMidEndSepPunct{\mcitedefaultmidpunct}
{\mcitedefaultendpunct}{\mcitedefaultseppunct}\relax
\EndOfBibitem
\bibitem[Wang et~al.(1980)Wang, Senbetu, and Woo]{Wang1980}
Wang,~S.; Senbetu,~L.; Woo,~C.-W. \emph{Journal of Low Temperature Physics}
  \textbf{1980}, \emph{41}, 611--628\relax
\mciteBstWouldAddEndPuncttrue
\mciteSetBstMidEndSepPunct{\mcitedefaultmidpunct}
{\mcitedefaultendpunct}{\mcitedefaultseppunct}\relax
\EndOfBibitem
\bibitem[Crowell and Brown(1982)]{crowell1982}
Crowell,~A.; Brown,~J. \emph{Surface Science} \textbf{1982}, \emph{123},
  296--304\relax
\mciteBstWouldAddEndPuncttrue
\mciteSetBstMidEndSepPunct{\mcitedefaultmidpunct}
{\mcitedefaultendpunct}{\mcitedefaultseppunct}\relax
\EndOfBibitem
\end{mcitethebibliography}

\end{document}